\documentclass[a4paper,10pt]{article}
\usepackage{enumerate}
\usepackage{color}
\usepackage[utf8]{inputenc} 
\usepackage[english]{babel}
\usepackage[T1]{fontenc}
\usepackage{graphicx}
\usepackage{amsfonts,amssymb,amsmath,latexsym,amsthm}
\usepackage{textcomp}
\usepackage[pdftex]{hyperref}
\usepackage{geometry}
\title{Interaction of interfacial waves with an external force: The Benjamin-Ono equation framework}
\author{Marcelo V. Flamarion$^{1}$ and Efim Pelinovsky$^{2,3}$}
\date{}

\begin{document}
\maketitle
\begin{center}
{\footnotesize $^1$Unidade Acad{\^ e}mica do Cabo de Santo Agostinho, \\
UFRPE/Rural Federal University of Pernambuco, BR 101 Sul, Cabo de Santo Agostinho-PE, Brazil,  54503-900 \\
marcelo.flamarion@ufrpe.br }

\vspace{0.3cm}
{\footnotesize $^{2}$Institute of Applied Physics, 46 Uljanov Str., Nizhny Novgorod 603155, Russia. \\
 $^{3}$Faculty of Informatics, Mathematics and Computer Science, HSE University, Nizhny Novgorod 603155, Russia. }


\end{center}


\begin{abstract} 
This study aims to explore the complex interactions between an internal solitary wave and an external force using the Benjamin-Ono equation as the theoretical framework. The investigation encompasses both asymptotic and numerical approaches. By assuming a small amplitude for the external force, we derive a dynamical system that describes the behavior of the solitary wave amplitude and the position of its crest. Our findings reveal three distinct scenarios: (i) resonance between the solitary wave and the external force, (ii) oscillatory motion with closed orbits, and (iii) displacement from the initial position while maintaining the wave direction. However, through numerical simulations, we observe a different relationship between the amplitude of the solitary wave and its crest position. Specifically, for external forces of small amplitude, the simulations indicate the presence of an unstable spiral pattern. Conversely, when subjected to external forces of larger amplitudes, the solitary wave exhibits a stable spiral trajectory which resembles the classical damped mass-spring system.
	\end{abstract}

\section{Introduction}
Considerable research efforts have been devoted to studying weakly nonlinear models that describe the evolution of internal waves. Prominent among these models are the Korteweg-de Vries (KdV) equation, which applies to shallow water, the Intermediate Long Wave (ILW) equation, suitable for fluids of finite depth, and the Benjamin-Ono (BO) equation, which pertains to deep water dynamics \cite{Benjamin:1967, Davis:1967, Ono:1975, Ko:1978}. These established models exhibit captivating characteristics, including the presence of periodic and solitary wave solutions that persist over time. However, it is essential to recognize that these model equations possess certain limitations that restrict their applicability to more generalized problems. Notably, they are valid only within specific depth ranges, thereby imposing a significant constraint on their practical utility.
\begin{figure}[!h]
	\centering	
	\includegraphics[scale =0.2]{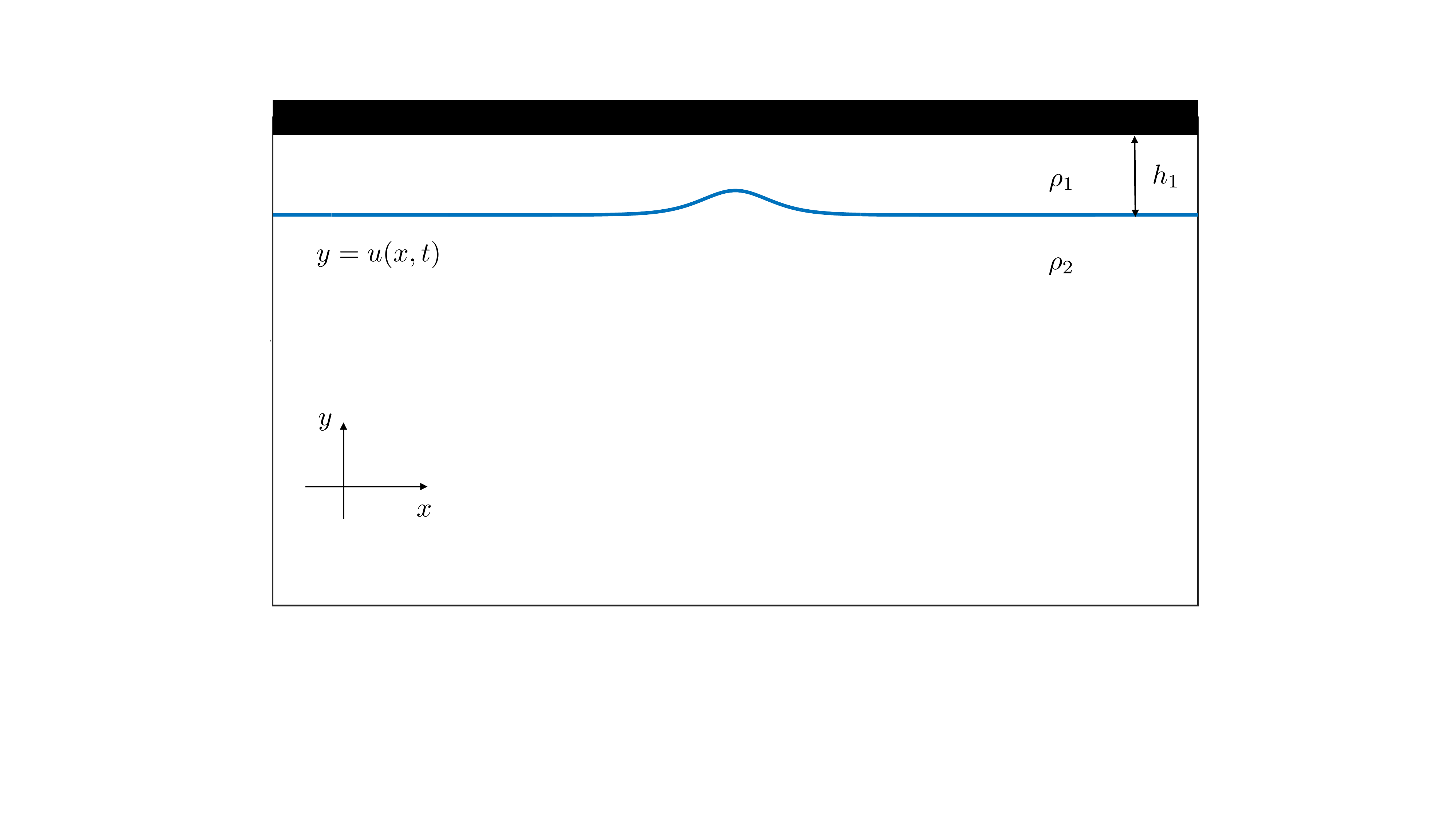}
	\caption{Sketch of the problem.}
	\label{Fig1}
\end{figure}

The renowned  Benjamin-Ono equation 
\begin{equation}\label{BO0}
u_{t} +c_{0}u_x-\frac{3c_0}{2h_1}uu_{x}+\frac{c_0h_1}{2\rho_r}\mathcal{H}[u_{xx}]=0,
\end{equation}
is commonly used to study the perturbed interface between two inviscid fluids of constant densities of a flat rigid lid and infinity depth. Here, $h_1$ is the thickness of the upper layer with density $\rho_1$, $\rho_2$ is the density of the lower fluid, $\rho_r=\rho_1/\rho_2<1$ is the ratio between the densities of the lighter fluid (upper layer) and the heavier one (lower layer) and $c_0$ is the linear speed given by
\begin{equation}\label{speed}
c_0^2=gh_{1}\Big(\frac{1}{\rho_r}-1\Big),
\end{equation}
where $g$ is the acceleration of gravity. More details of the geometry of the problem is depicted in Figure \ref{Fig1}. The elevation of the interface in the position $x$ and time $t$ is denote by $u(x,t)$ and $\mathcal{H}$ denotes the Hilbert transform defined as
\begin{equation}\label{Hilbert}
\mathcal{H}[u(x,t)]=\int_{-\infty}^{+\infty}\frac{u(y,t)}{y-x}dy.
\end{equation}

\begin{figure}[!h]
	\centering	
	\includegraphics[scale =0.3]{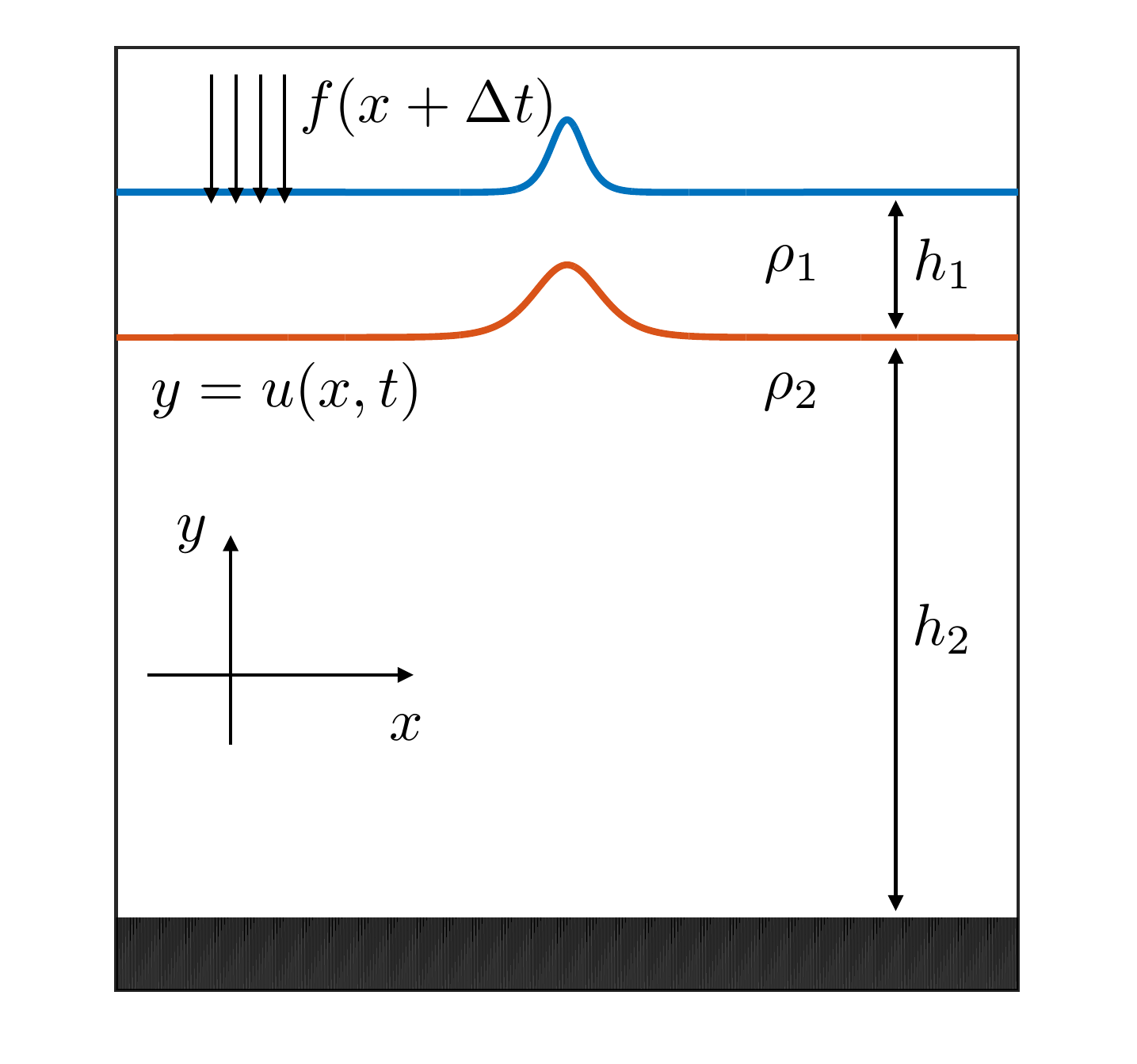}
		\includegraphics[scale =0.3]{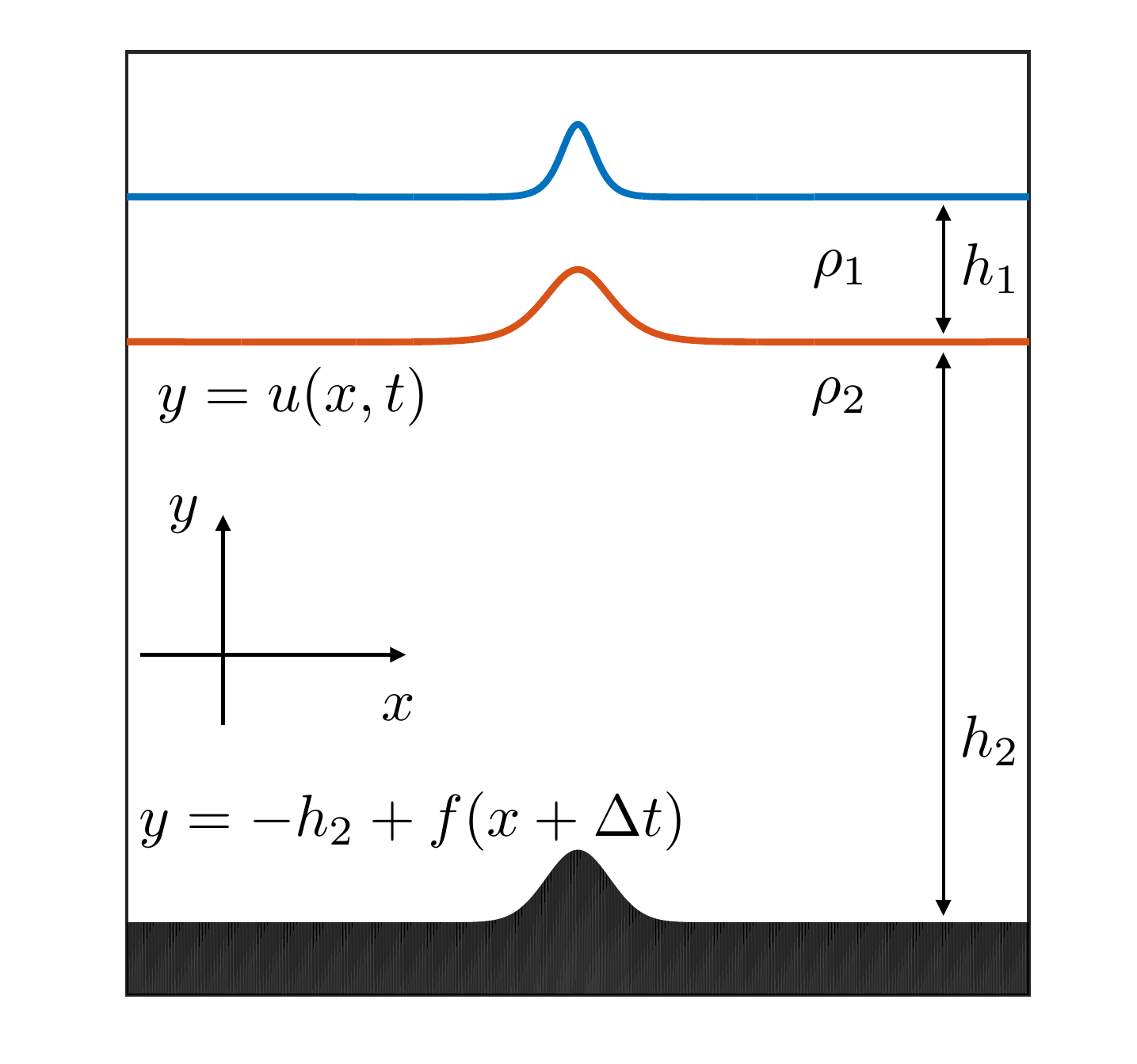}
	\caption{Sketch of the problem. Left: A moving with constant speed ($\Delta$) disturbance at the free surface. Right: A moving obstacle with constant speed ($\Delta$) at the bottom.}
	\label{Fig1b}
\end{figure}

One of the key issues in water wave research is the investigation of the interaction between solitary waves and a heterogeneous medium. Various frameworks have been employed to study this problem. Integrable models, such as the Korteweg-de Vries (KdV) and modified KdV (mKdV) equations, have been explored extensively \cite{Baines, Ermakov, Lee, LeeWhang, Kim, COAM, Collisions, Capillary2, Malomed:1993, Grimshaw:1993, Wu1, Paul, Grimshaw86, Chaos:FP, Flamarion-Pelinovsky:2022a}. Additionally, nonintegrable models including the Whitham \cite{Whitham1, Whitham2}, Schamel \cite{Chowdhury:2018}, and Euler equations \cite{COAM2} have been utilized . However, to the best of our knowledge, the study of this phenomenon within the framework of the Benjamin-Ono (BO) equation has not been addressed in the existing literature. The inclusion of an external force in equation (\ref{BO0}) introduces intriguing physics problems. For instance, the external force commonly arises in two scenarios: (i) a pressure distribution is applied at the free surface of the upper layer \cite{Choi:1996, Smyth:2002}, and (ii) the external force can represent bathymetry \cite{Matsuno:1993a, Matsuno:1993b, Choi:1996}. The latter case can effectively model flow over a mountain in the atmosphere when the upper layer extends to infinity. These two cases are depicted in Figure \ref{Fig1b}.

The aim of this this study is to explore the interaction between an internal solitary wave and an external force. To accomplish this, we investigate the forced Benjamin-Ono equation. By assuming a small amplitude for the external force, we derive a two-dimensional dynamical system that characterizes the position of the solitary wave crest and its amplitude. We then compare asymptotic results with fully numerical simulations conducted using pseudo-spectral methods. Our findings indicate that, at early times, the results exhibit qualitative agreement. However, the asymptotic theory predicts the presence of centers and closed orbits when the external force and the solitary wave approach resonance. In contrast, the fully numerical simulations suggest the occurrence of unstable spirals.

For reference, this article is organized as follows: The forced BO equation is presented in Section 2. In Section 3, we describe the asymptotic and numerical results. Then, we present the final considerations in Section 4.

\section{The forced BO equation}
Our focus of study lies in exploring the interaction between internal solitary waves and an external force field. To accomplish this, we examine the Benjamin-Ono (BO) equation in its canonical form, incorporating an external force function $f(x)$ and a constant speed $\Delta$
\begin{equation}\label{BO1}
u_{t} +uu_{x}+\mathcal{H}[u_{xx}]=f_{x}\Big(x+\Delta t\Big).
\end{equation}
This equation encompasses terms such as the convective nonlinearity, dispersion through the Hilbert transform $\mathcal{H}$, and the external force acting on the wave field represented by $u(x,t)$. 

Our objective is to delve into the intricate dynamics of solitary waves when subjected to the influence of this external force field. For convenience, we transform Equation (\ref{BO1}) into the moving frame associated with the external force. This transformation is achieved by introducing the new variables $x' = x + \Delta t$ and $t' = t$. Within this new coordinate system, Equation (\ref{BO1}) can be expressed as
\begin{equation}\label{BO2}
u_{t} +\Delta u_{x}+uu_{x}+\mathcal{H}[u_{xx}]=f_{x}(x),
\end{equation}
wherein the effect of the external force is accounted for solely in terms of its spatial derivative $f_x(x)$.

In this context, it is crucial to note that the mass of the system, represented by the integral of the wave field over space, is preserved. This conservation of mass is mathematically expressed by
\begin{equation}\label{mass}
M(t) = \int_{-\infty}^{+\infty}u(x,t) dx.
\end{equation}
Furthermore, the momentum of the system, denoted as $P(t)$, is balanced by the external force as
\begin{equation}\label{momentum0}
\frac{dP}{dt} = \int_{-\infty}^{+\infty}u(x,t)f_{x}(x) dx, \mbox{ where } P(t)=\int_{-\infty}^{+\infty}u^{2}(x,t)dx.
\end{equation}
These mass and momentum formulas, as stated in equations (\ref{mass})-(\ref{momentum0}), hold significant importance, particularly in evaluating the accuracy and dependability of numerical methods employed for solving the BO equation (\ref{BO2}). By utilizing these formulas, one can assess the precision of the numerical methodologies utilized and gain confidence in the reliability of the obtained results.

When there is no external force present, the BO equation (\ref{BO2}) admits a two-parameter $(a,\lambda)$ family of periodic waves as solutions \cite{Benjamin:1967}. These waves can be described by the following expressions \cite{Benjamin:1967}
\begin{equation}\label{periodic}
u(x,t)=\frac{A}{1-B\cos\Big(\frac{2\pi}{\lambda}(x-ct)\Big)}, \mbox{  where }  \; c =\Delta+ \frac{a}{4}, \; A=\frac{32\pi^2}{a\lambda^{2}} \mbox{ and } B = \Bigg[1-\Big(\frac{8\pi}{a\lambda}\Big)^{2}\Bigg]^{1/2}.
\end{equation}
As $\lambda\rightarrow\infty$, it reduces to the solitary wave solution  \cite{Benjamin:1967}
\begin{equation}\label{solitary}
u(x,t)=\frac{al^2}{(x-ct)^2+l^2}, \mbox{  where }  \; c =\Delta+\frac{a}{4} \mbox{ and } |l| = \frac{4}{a}.
\end{equation}
Here, $a$ represents the solitary wave amplitude, $c$ represents its speed, and $l$ characterizes the solitary wavenumber.

In our investigation, we consider the external force to be of the form
\begin{equation}\label{externalforce}
f(x) = b \exp\Big(-\frac{x^{2}}{w^2}\Big),
\end{equation}
where $b$ is the force amplitude and $w$ is its width and focus on the interaction of the solitary waves  (\ref{solitary}) and the external force (\ref{externalforce}).

\section{Results}
\subsection{Asymptotic results}
In this section, our objective is to deduce the governing equations for the interaction between solitary waves and an external force, considering the force to have a small amplitude. To accomplish this, we introduce a small positive parameter, denoted as $\epsilon$, and substitute $\epsilon f$ for the external force $f$ in equation (\ref{BO2}). Moreover, we assume that the wave field closely resembles a solitary wave, characterized by parameters that slowly vary over time \cite{Grimshaw94, Grimshaw2002, Pelinovsky:2002}. Mathematically, the solitary wave can be described using the following expressions
\begin{equation}\label{solitary}
u(\mathbf{\Phi},T)=\frac{a(T)l(T)^2}{\mathbf{\Phi}^2+l(T)^2}, \mbox{  where }  \;\ \mathbf{\Phi}=x-X(T) \mbox{ and } X(T) =x_0+\frac{1}{\epsilon}\int_{0}^{T} c(T)dT,
\end{equation}
where the solitary wave  initial position is denoted as $x_0$, and the functions $a$ and $c$ are established based on the interaction between the wave field and the external field. To facilitate our analysis, we introduce the concept of slow time by introducing a new variable, namely $T=\epsilon t$. We aim to find a solution by utilizing an asymptotic expansion in the following form
\begin{align} \label{Asymptotic}
\begin{split}
& u(\mathbf{\Phi},T)=u_{0}+\epsilon u_1+\epsilon^{2} u_2+\cdots , \\
& c(T) = c_0 + \epsilon c_1 + \epsilon^{2} c_2+\cdots. \\
\end{split}
\end{align}
At the lowest order of the perturbation theory, it immediately follows that the solutions $u_0$ and $q_0$ are precisely defined in accordance with equation (\ref{solitary}).

The  momentum balance equation at the first-order is 
\begin{equation}\label{momentum}
\frac{1}{2}\frac{d}{dT}\int_{-\infty}^{\infty} u_{0}^{2}(\mathbf{\Phi})d\mathbf{\Phi} = \epsilon\int_{-\infty}^{\infty}u_{0}(\mathbf{\Phi})\frac{df}{d\mathbf{\Phi}}(\mathbf{\Phi}+X)d\mathbf{\Phi}. 
\end{equation}
Therefore, replacing the formulas (\ref{solitary}) into  the equation (\ref{momentum}) yields the two-dimensional dynamical system
\begin{align} \label{DS0}
\begin{split}
& \frac{da}{dt} = \int_{-\infty}^{\infty}\Bigg[\frac{al^2}{\mathbf{\Phi}^2+l^2}\Bigg]\frac{df}{d\mathbf{\Phi}}(\mathbf{\Phi}+X)d\mathbf{\Phi}, \\
& \frac{dX}{dt}=\Delta+\frac{a}{4}.
\end{split}
\end{align}

When the external force extends far beyond the scope of the solitary wave wavelength, it becomes feasible to approximate the solitary waves as delta functions. This approximation allows us to simplify the dynamical system that governs the amplitude and position of the crest of the solitary wave. In doing so, we arrive at the following simplified form, which encapsulates the essence of the solitary wave behavior
\begin{align} \label{DS}
\begin{split}
& \frac{da}{dt}=\frac{df}{dX}(X), \\
& \frac{dX}{dt} = \Delta+\frac{a}{4}.
\end{split}
\end{align}
From equation (\ref{DS}) we have that the position of the crest of a solitary wave is described by the oscillator
\begin{equation}
\frac{d^{2}X}{dt^{2}}=\frac{1}{4}f(X).
\end{equation}
This is similar to what happens to the forced KdV equation \cite{Grimshaw94, Grimshaw96} and the forced mKdV \cite{Pelinovsky:2002, Flamarion-Pelinovsky:2022a}.

Equilibrium points in the dynamical system (\ref{DS}) exist exclusively when $\Delta$ assumes a negative value. The magnitude of the solitary wave amplitude ($a_0$) and the location of its peak ($X_0$) are
\begin{equation} \label{equilibrium}
a_{0}=-\frac{a}{4} \mbox{ and } X_{0}=0 \mbox{ for } \Delta<0.
\end{equation}
When the disturbance and the solitary wave exhibit the same polarity, the equilibrium position is categorized as a center. Centers correspond to stable solitary waves that remain steady. The interaction between the solitons and the external force concludes once the waves have traversed the external force region.
\begin{figure}[!h]
	\centering	
	\includegraphics[scale =1]{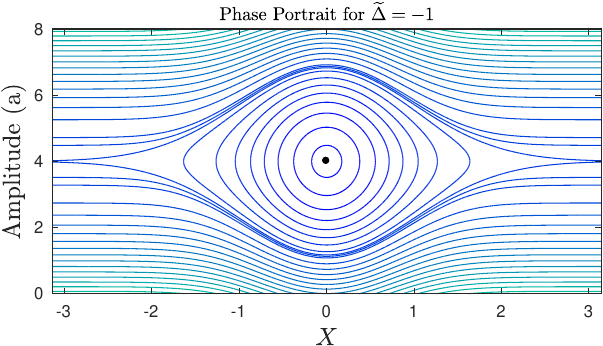}
	\caption{Phase portraits for the dynamical system (\ref{DS}). Dot corresponds to the equilibrium point.}
	\label{Fig2}
\end{figure}

The streamlines, which are the level curves of the stream function, serve as a representation of the solutions of the system (\ref{DS}). They are represented by the Hamiltonian
\begin{equation}\label{streamfunction}
\mathbf{\Psi}(X,a) = -f(X)+\Delta a+\frac{a^2}{8}.
\end{equation}

To examine the phase portrait of system (\ref{DS}) with the external force (\ref{externalforce}), we introduce a rescaling of variables. The coordinate $X$ is scaled relative to $w$, while the amplitude $a$ is scaled relative to $b^{1/2}$. Here, $b>0$ represents the amplitude of the external forcing. This rescaling results in the emergence of a new parameter 
\begin{equation}
\widetilde{\Delta}= \frac{\Delta}{\sqrt{b}}.
\end{equation}
With these new scalings, the  stream function becomes
\begin{equation}\label{streamfunction}
\mathbf{\Psi}(X,a) = -e^{-{X^2}}+\tilde{\Delta}a+\frac{a^2}{8}.
\end{equation}

Figure \ref{Fig1} illustrates typical phase portraits of system (\ref{DS}). It is crucial to emphasize that the presence of a closed orbit in the phase portrait signifies the existence of a solitary wave. This solitary wave is effectively confined without any radiation, which arises due to its interaction with the external force.

\subsection{Numerical results}
The numerical solution of equation (\ref{BO2}) is obtained by employing a Fourier pseudospectral method with an integrating factor. The equation is solved in a periodic computational domain $[-L, L]$ with a uniform grid containing $N$ points. This grid allows us to approximate the spatial derivatives accurately \cite{Trefethen:2000}.  To mitigate the influence of spatial periodicity, the computational domain is chosen to be sufficiently large. For the time evolution of the equation, we employ the classical fourth-order Runge-Kutta method with discrete time steps of size $\Delta t$. The external force is chosen as in equation (\ref{externalforce}) with $b=1$ and $w=10$. For a  complete study resolution of a similar numerical method, the readers are refer to the work of Flamarion et al. \cite{Marcelo-Paul-Andre}. 
\begin{figure}[!h]
	\centering	
	\includegraphics[scale =0.97]{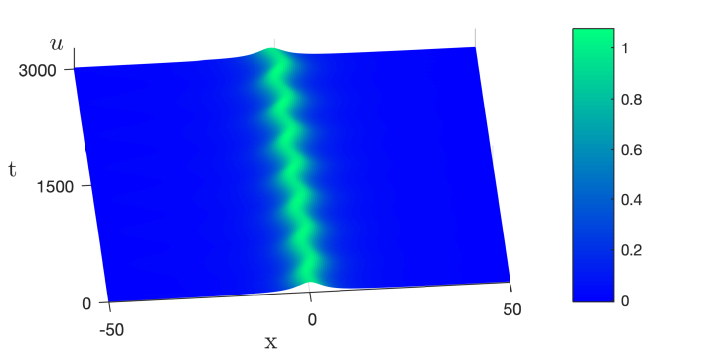}
	\includegraphics[scale =0.97]{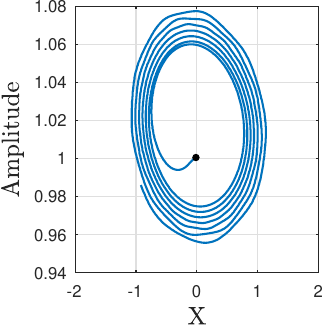}
	\includegraphics[scale =0.97]{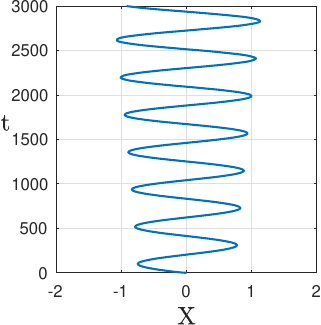}
	\caption{Top: trapped solitary wave at the external force. Bottom (left):
the solitary wave amplitude vs. crest position. Bottom (right): the crest position along the time.
Parameters: $a=1$, $w=10$, $\Delta=-0.25$ and $\epsilon=0.01$.}
	\label{Fig3}
\end{figure}

To verify the asymptotic results described in the previous section, we conduct a series of simulations using the BO equation (\ref{BO2}). We set the initial amplitude of the solitary wave to $a=1$ and the speed deviation $\Delta=-a/4$. With these parameters, the dynamical system (\ref{DS}) predicts steady solutions, and small perturbations of these values result in closed orbits or trapped waves without radiation.

Initially, we consider $\epsilon=0.01$. In this case, we observe that the solitary wave oscillates back and forth over the external force for extended periods, as shown in Figure \ref{Fig3} (Top). The fluctuation in the amplitude was found to be of the order of $\mathcal{O}(10^{-2})$, indicating that the amplitude of the solitary wave remains nearly unchanged over time. Although the results closely match the asymptotic theory predictions at small times, the fully numerical computations revealed a behavior resembling an unstable spiral in the amplitude vs. crest position space and a resonant harmonic oscillator in the crest position vs. time space, as illustrated in Figure \ref{Fig3} (Bottom). Notably, minimal radiation was observed during the interaction between the solitary wave and the external force, as shown in Figure \ref{Fig5}.
\begin{figure}[!h]
	\centering	
	\includegraphics[scale =1]{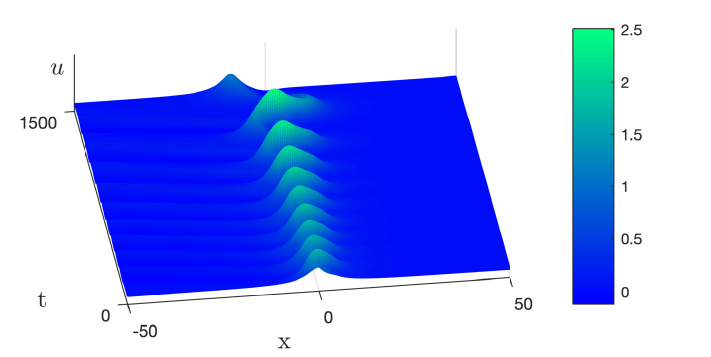}
	\includegraphics[scale =1]{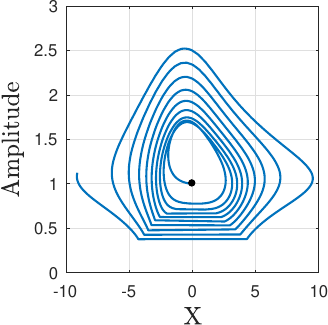}
	\includegraphics[scale =1]{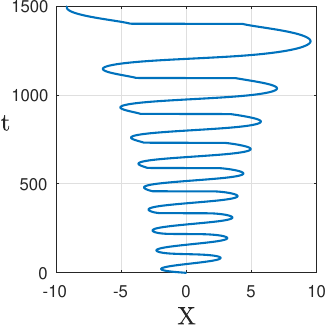}
	\caption{Top: trapped solitary wave at the external force. Bottom (left):
the solitary wave amplitude vs. crest position. Bottom (right): the crest position along the time.
Parameters: $a=1$, $w=10$, $\Delta=-0.25$ and $\epsilon=0.1$.}
	\label{Fig4}
\end{figure}

Next, we increase the value of the parameter to $\epsilon =0.1$. In this case, the solitary wave remains trapped at the external force for long durations, as depicted in Figure \ref{Fig4} (Top). However, there were significant differences compared to the previous case. The amplitude oscillations were much larger, as shown in Figure \ref{Fig4} (Bottom-left). Moreover, the oscillations in the crest position vs. time space were more pronounced, see \ref{Fig4} (Bottom-right). Nevertheless, the solitary wave continue to remain trapped at the external force.
\begin{figure}[!h]
	\centering	
	\includegraphics[scale =1]{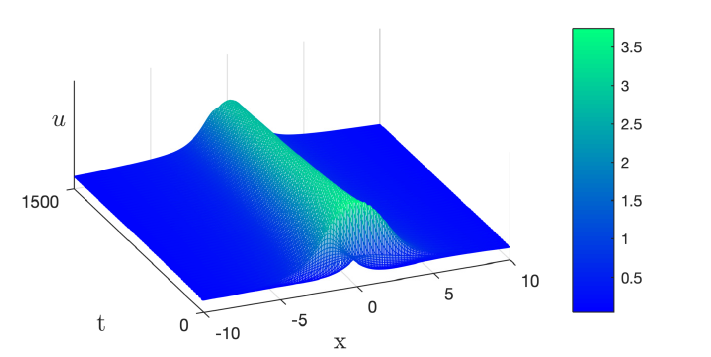}
	\includegraphics[scale =1]{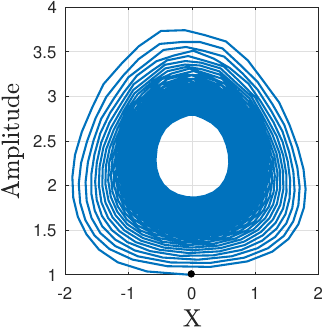}
	\includegraphics[scale =1]{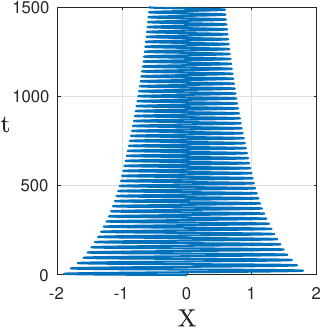}
	\caption{Top: trapped solitary wave at the external force. Bottom (left):
the solitary wave amplitude vs. crest position. Bottom (right): the crest position along the time.
Parameters: $a=1$, $w=10$, $\Delta=-0.25$ and $\epsilon=0.5$.}
	\label{Fig5}
\end{figure}\

\begin{figure}[!h]
	\centering	
	\includegraphics[scale =1]{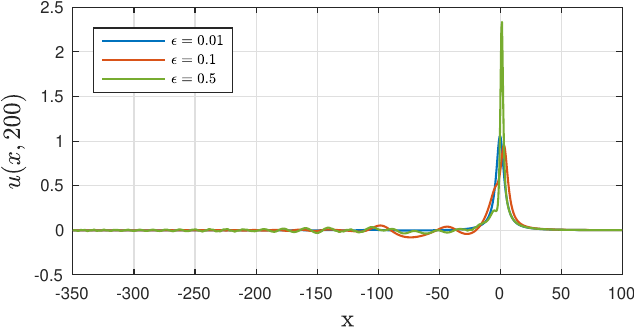}
	\includegraphics[scale =1]{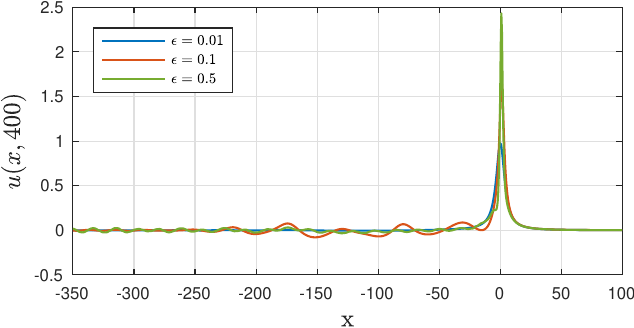}
	\caption{Comparison between the trapped waves for different values of the parameter $\epsilon$ at different times.
Parameters: $a=1$, $w=10$, $\Delta=-0.25$.}
	\label{Fig6}
\end{figure}

Lastly, we further increase the parameter to $\epsilon =0.5$. In this scenario, we expect an increase in radiation and the possibility of the solitary wave moving past the external force at earlier times. However, the increased value of $\epsilon$ causes a substantial increase in the amplitude of the solitary wave, as depicted in Figure \ref{Fig5} (Top). The resulting amplitude of the solitary wave becomes much larger than that of the external force, effectively rendering the presence of the external force negligible in the dynamics. The fully numerical computations revealed a behavior resembling a stable spiral in the amplitude vs. crest position space and a damped harmonic oscillator in the crest position vs. time space, as shown in Figure \ref{Fig5} (Bottom). Furthermore, minimal radiation was observed in the interaction between the solitary wave and the external force, as depicted in Figure \ref{Fig6}.

\section{Conclusion}
In this work, our focus was to examine the interactions between internal solitary waves and an external force. Utilizing asymptotic expansion techniques, we derived a simplified model that describes the position and amplitude of the solitary waves. However, when comparing our numerical results with the asymptotic model, we found agreement only in the early stages. Interestingly, while the asymptotic predictions indicated steady solutions and perfect trapping, the fully numerical solutions showed the emergence of unstable and stable spirals, depending on the magnitude of the external force. This discrepancy underscores the significance of considering higher-order terms in the asymptotic expansion. We believe that incorporating these higher-order terms will enable the asymptotic theory to not only predict the positions of solitary waves but also capture the occurrence of unstable spirals, which closely resemble the outcomes of the numerical simulations. Therefore, a logical next step would involve further investigation and comparison of numerical and asymptotic solutions, incorporating higher-order terms, as part of our future research direction.

\section{Acknowledgements}
M.V.F is grateful to IMPA for hosting him as visitor during the 2023 Post-Doctoral Summer Program. E.P. is supported by support by the RNF grant number 19-12-00253

	\section*{Declarations}
	
	\subsection*{Conflict of interest}
	The authors state that there is no conflict of interest. 
	\subsection*{Data availability}
	
	Data sharing is not applicable to this article as all parameters used in the numerical experiments are informed in this paper.


\begin{thebibliography}{999}
	
				\bibitem{Baines} {Baines, S.} 
      { Topographic effects in stratified flows.}
	{\it  Cambridge University Press, Cambridge.} {\bf 1995}. 
	

			\bibitem{Benjamin:1967}
	{Benjamin T.B.} 
	Internal waves of permanent form of great depth.
	{\it J Fluid. Mech.}  {\bf 1967}, {295}, 381-394.	
	
		\bibitem{Choi:1996}
	{Choi, W.; Camassa, R.} 
	Weakly nonlinear internal waves in a two-fluid system.
	{\it J Fluid. Mech.}  {\bf 1996}, {313}, 83-103.	
	
		\bibitem{Chowdhury:2018}{ Chowdhury, S.; Mandi, L.;  Chatterjee, P. } 
	{Effect of externally applied periodic force on ion acoustic waves
in superthermal plasmas.}
	{\it Phys. of Plasma.} {\bf 2018} 25, 042112. 	
	
			\bibitem{Davis:1967}
	{Davis R.E.; Acrivos A.} 
	Solitary internal waves in deep water.
	{\it J Fluid. Mech.}  {\bf 1967}, {295}, 593-607.			
	

	
		
			\bibitem{Ermakov}
	{Ermakov A.; Stepanyants, Y.} 
	Soliton interaction with external forcing within the Korteweg-de Vries equation
	{\it Chaos.}  {\bf 2019}, {29}, 013117.
	
		\bibitem{Marcelo-Paul-Andre}{Flamarion MV, Milewski PA,  Nachbin A.} 
				{Rotational waves generated by current-topography interaction.}
				{\it Stud Appl Math.} 2019;142:433-464. 
	
			\bibitem{Collisions}{Flamarion,  M.V.; Ribeiro-Jr, R.} 
		    {Solitary water wave interactions for the Forced Korteweg-de Vries equation.}
	{\it Comp. Appl. Math.} {\bf 2021}, 40, 312.
	
	
			\bibitem{Capillary2}{Flamarion, M.V.; Ribeiro-Jr, R.} 
        {Gravity-capillary flows over obstacles for the fifth-order forced Korteweg-de Vries equation.}
	{\it J. Eng. Math.} {\bf 2021}, 129, 1-17.

	
	
	
	\bibitem{COAM}{Flamarion, M.V.} 
		    {Generation of trapped depression solitary waves in gravity-capillary flows over an obstacle.}
	{\it Comp. Appl. Math.} {\bf 2022}, 41, 31.
	
		\bibitem{COAM2}{Flamarion, M.V.; Ribeiro-Jr, R.} 
		    {Trapped solitary-wave interaction for Euler equations with low-pressure region.}
	{\it Comp. Appl. Math.} {\bf 2021}, 40, 20.
	
			\bibitem{Chaos:FP}{Flamarion, M.V.; Pelinovsky E.} 
		    {Soliton interactions with an external forcing: the modified Korteweg-de Vries framework.}
	{\it Chaos, Solitons \& Fractals.} {\bf 2022}, 165, 112889.
	
				\bibitem{Whitham1}{Flamarion, M.V.} 
        {Waves generated by a submerged topography for the Whitham equation.}
	{\it Int. J. Appl. Comput. Math.} {\bf 2022}, 8, 257.
	
					\bibitem{Whitham2}{Flamarion, M.V.} 
        {Trapped waves generated by an accelerated moving disturbance for the
Whitham equation.}
	{\it Partial Differential Equations in Applied Mathematics.} {\bf 2022}, 5, 100356.
	

   \bibitem{Flamarion-Pelinovsky:2022a}{Flamarion MV, Pelinovsky E.} 
	{Solitary wave interactions with an external periodic force: The extended Korteweg-de Vries framework.}
	{\it Mathematics.} {\bf 2022}, 10, 4538.
	
	
	
	

	
	


	\bibitem{Grimshaw94}
	{Grimshaw, R.; Pelinovsky,  E.;  Tian, X.} 
	Interaction of a solitary wave with an external force.
	{\it Physica D.}  {\bf 1994}, 77, 405-433.
	

	
	\bibitem{Grimshaw96}
	{Grimshaw R.; Pelinovsky, E.; Pavel, S.} 
	Interaction of a solitary wave with an external force moving with variable speed.
	{\it Stud. Appl. Math.}  {\bf 1996}, {142}, 433-464.
	
			

	
			\bibitem{Grimshaw:1993}
	{Grimshaw, R.; Malomed, B.A.; Tian, X.} 
	Dynamics of a KdV soliton due to periodic forcing.
	{\it Phys. Lett. A.}  {\bf 1993}, {179}, 291-298.
			
	
		\bibitem{Grimshaw2002}
	{Grimshaw, R.; Pelinovsky, E.} (2002)
	Interaction of a solitary wave with an external force in the extended Korteweg-de Vries equation.
	{\it Int. J. Bifurcat. Chaos.}  {\bf 2002}, {12}(11), 2409-2419.
	
	
	
	
	

	

%

%
		      \bibitem{Grimshaw86}{Grimshaw R,  Smyth N.} 
{Resonant flow of a stratified fluid over topography in water of finite depth.}
	{\it J. Fluid Mech.} {\bf 1986}, 169, 235-276. 
%


			\bibitem{Kim} {Kim, H.; Choi, H.} 
      {A study of wave trapping between two obstacles in the forced Korteweg-de Vries equation.}
	{\it J. Eng. Math.} {\bf 2018}, 108, 197-208. 
	
		\bibitem{Ko:1978} {Kubota, T.; Ko D.R.S.; Dobbs, L.D.} 
      {Propagation of weakly nonlinear internal waves in a stratified fluid of finite depth.}
	{\it AZAA J. Hydrodyn.} {\bf 1978}, 12, 157-165.


	\bibitem{Lee} {Lee, S.} 
      {Dynamics of trapped solitary waves for the forced KdV equation.}
	{\it Symmetry.} {\bf 2018}, 10(5), 129. 

\bibitem{LeeWhang} {Lee, S.;  Whang, S.} 
      { Trapped supercritical waves for the forced KdV equation with two bumps.}
	{\it Appl. Math. Model.} {\bf 2015}, 39, 2649-2660. 
	
					\bibitem{Malomed:1993}{Malomed, B.A.} 
		    {Emission of radiation by a KdV soliton in a periodic forcing.}
	{\it Phys. Lett. A.} {\bf 1993}, 172, 373-377.
	
		\bibitem{Matsuno:1993a}
	{Matsuno, Y.} 
	A unified theory of nonlinear wave propagation in two-layer fluid systems.
	{\it Phys. Soc. Japan}  {\bf 1993}, {62}, 1902-1916.	
	
	\bibitem{Matsuno:1993b}
	{Matsuno, Y.} 
	Nonlinear evolution of surface gravity waves over an uneven bottom.
	{\it J. Fluid. Mech}  {\bf 1993}, {249}, 121-133.	
	
			\bibitem{Ono:1975}
	{Ono H.} 
	Algebraic solitary waves in stratified fluids.
	{\it J Phys. Soc. Japan}  {\bf 1975}, {39}, 1082-1091.	
	
			\bibitem{Paul}{Milewski PA. } 
	{The Forced Korteweg-de Vries equation as a model for waves generated by topography.}
	{\it Cubo Math J.} {\bf 2004}, 6, 33-51. 

%

	
%

	
		\bibitem{Pelinovsky:2002}
	{Pelinovsky E.} (2002)
	Autoresonance processes under interaction of solitary waves with
the external fields.
	{\it Int. J Fluid. Mech. Res.}  {\bf 2002}, {30}(5), 493-501.
	



		\bibitem{Smyth:2002}{Porter, A.; Smyth, N.} 
        {Modelling the morning glory of the Gulf of Carpentaria.}
	{\it J Fluid Mech.} {\bf 2002}, 454, 1-20. 


	
		\bibitem{Trefethen:2000}{Trefethen, L.N.} 
        {\it Spectral Methods in MATLAB.}
	{Philadelphia: SIAM;} 2001.
	
	

	
		\bibitem{Wu1}{Wu TY.} 
        {Generation of upstream advancing solitons by moving disturbances.}
	{\it J Fluid Mech.} {\bf 1987}, 184, 75-99. 
	
	
%
%
%
	
	




		
%
%

         
	

	
	
%



	


%


	

		
		
		
	\end{thebibliography}
\end{document}